# NEURAL NETWORK MODEL OF THE PXIE RFQ COOLING SYSTEM AND RESONANT FREQUENCY RESPONSE*

A.L. Edelen†, S.G. Biedron[1], S.V. Milton, Colorado State University, Fort Collins, CO
D. Bowring, B.E. Chase, J.P. Edelen, J. Steimel, Fermilab, Batavia, IL
[1]also at University of Ljubljana, Ljubljana, Slovenia

*Abstract*

As part of the PIP-II Injector Experiment (PXIE) accelerator, a four-vane radio frequency quadrupole (RFQ) accelerates a 30-keV, 1-mA to 10-mA H$^-$ ion beam to 2.1 MeV. It is designed to operate at a frequency of 162.5 MHz with arbitrary duty factor, including continuous wave (CW) mode. The resonant frequency is controlled solely by a water-cooling system. We present an initial neural network model of the RFQ frequency response to changes in the cooling system and RF power conditions during pulsed operation. A neural network model will be used in a model predictive control scheme to regulate the resonant frequency of the RFQ.

## INTRODUCTION AND MOTIVATION

One of the main challenges for the PXIE RFQ operation is to maintain the specified field across the vane tips for beam acceleration. The RF amplifiers have enough power capability to maintain the field when the RFQ is out of tune by up to 3 kHz.

As the PXIE RFQ is expected to operate in both CW and pulsed RF mode with a wide range of duty factors, the expected RF heating will thus vary significantly, resulting in variable detuning of the cavity. The resonant frequency is particularly sensitive to changes in the vane pole tip positions resulting from both local and bulk heating/cooling (i.e. in the vanes and walls respectively). Because of their lower thermal mass, the vanes also contract and expand faster than the walls given the same heating/cooling power. Together, this results in a large transient response in resonant frequency to changes in average RF power or water temperature disturbances (e.g. see Figure 2). Transport delays, thermal time constants, and coupling between the wall and vane circuits further complicate the control task.

Thermal expansion and contraction of the RFQ vanes and walls will be leveraged to ensure that the desired resonant frequency is maintained in spite of RF heating. In simulations conducted at LBNL, the estimated frequency response is -16.7 kHz/°C in the vanes and 13.9 kHz/°C in the walls [1, 2]. Independent control of the wall and vane circuits enables exploitation of the individual frequency responses to provide a wider tuning range.

In light of these challenges, model predictive control (MPC) will be used for resonance control. The controller will use measurements from the water system and RF system to plan future sequences of the vane and wall valve settings to adjust the water temperatures.

As part of this effort, an initial neural network model of the system was created. For a discussion of neural networks for particle accelerator modeling and control, see [3]; this also contains an example of MPC applied to a similar accelerator subsystem. The RFQ is described in further detail in [1, 2], and [4] provides an overview of the PXIE RFQ resonance control effort.

## RFQ COOLING SYSTEM

A simplified diagram of the RFQ cooling system is shown in Figure 1. The system consists of two water sub-circuits that supply temperature-regulated water to the vane channels and the wall channels machined into the RFQ. These sub-circuits are driven by a cooling skid connected to the laboratory's low conductivity water (LCW) supply. It is important to note that the vane and wall sub-circuits are coupled through pressure balancing in the system, mixing in the skid return line, and heat flow through the RFQ itself. A portion of the warm water from the exit of the RFQ is returned to each sub-circuit supply.

Two flow control valves on each sub-circuit determine the ratio of cool to warm water that is mixed together prior to entering the RFQ. In-line helical mixers ensure that the mixing is thorough. Note that for each of the four individual modules comprising the RFQ, there are eight wall cooling channels and four vane cooling channels.

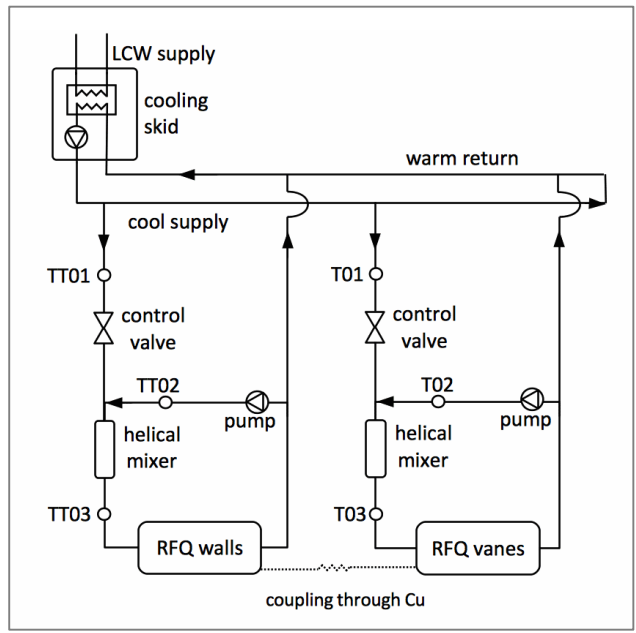

Figure 1: Simplified PXIE RFQ cooling system diagram. TT01, T01, et cetera are temperature sensors.

---

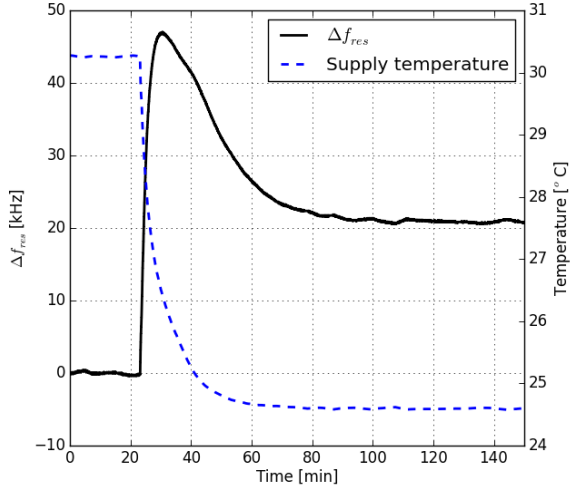

Figure 2: Measured, uncontrolled change in resonant frequency after a roughly 5 °C reduction in the cold supply temperature from the cooling skid.

## MEASURED DATA

Characterization data were obtained from the RFQ during pulsed operation. Parameter scans yielded the frequency response to various combinations of vane valve settings, wall valve settings, and RF field amplitudes. Figure 3 shows a representative scan over the vane valve setting and RF field amplitude.

In total, the wall valve was varied from 0% to 99% open, and the vane valve was varied from 0% to 99% open. The cavity field was varied from 0 kV to 70 kV. For all data sets, the repetition rate was 10 Hz and the pulse duration was 4 ms. LLRF feedback was used during the test to regulate the cavity field and maintain the drive frequency at 162.4650 MHz. Table 1 shows the range of measured system parameters.

For these measurements, the resonant frequency is calculated from the RF signals as follows:

$$\phi_f = \operatorname{atan}\left(\frac{E_{f1}\sin(\phi_{f1}) + E_{f2}\sin(\phi_{f2})}{E_{f1}\cos(\phi_{f1}) + E_{f2}\cos(\phi_{f2})}\right)$$

$$\delta = \frac{\tan(\phi_c - \phi_f)}{2Q_L} f_0$$

where $E_{f1}$ and $E_{f2}$ are the magnitudes of the drive signals from each amplifier, $\phi_{f1}$ and $\phi_{f2}$ are the forward phases of each drive signal, $\phi_f$ is the calculated forward phase, $\phi_c$ is the cavity phase, $f_0$ is the drive frequency, $Q_L$ is the loaded quality factor of the cavity, and $\delta$ is the detuning (note that this does not take into account beam loading). Because the RFQ is driven by two amplifiers, we need to calculate the vector sum of the two forward signals in order to obtain the forward phase for the calculation of resonant frequency shift. Over the training data, the resonant frequency ranges from 162.4403 MHz to 162.4738 MHz.

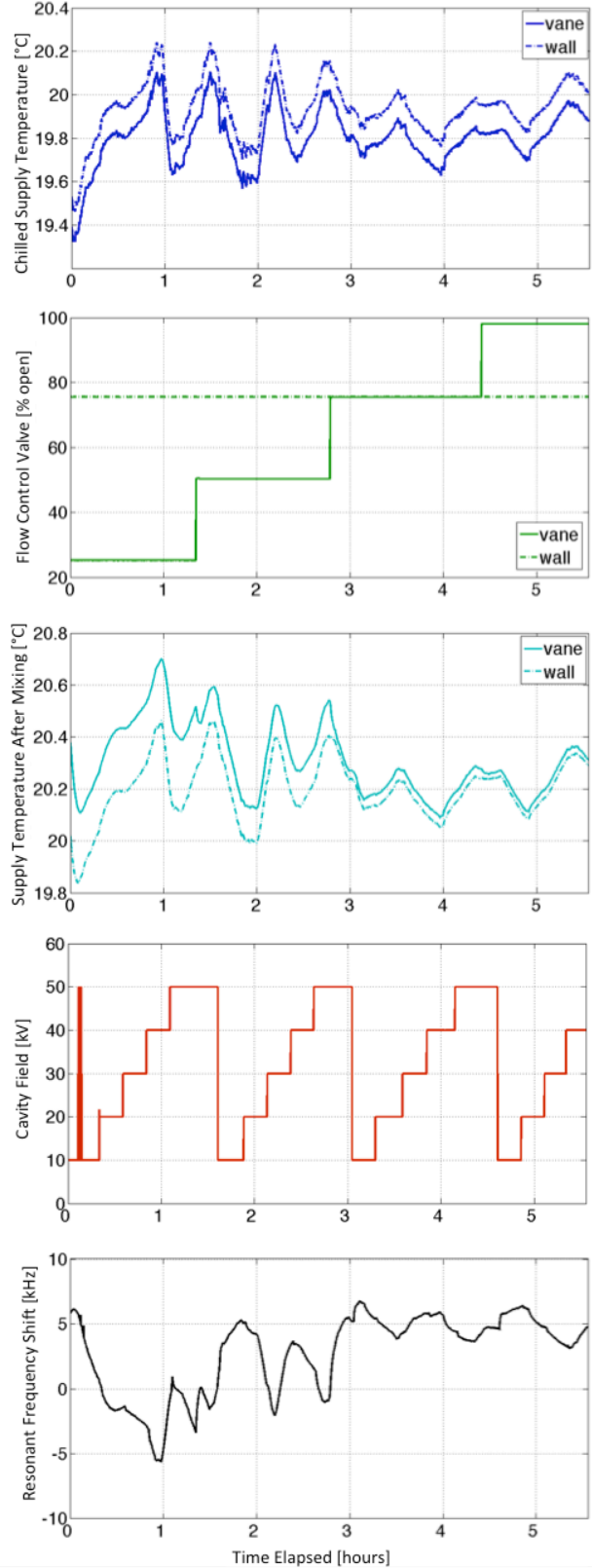

Figure 3: An example scan from the measured data. The repetition rate was 10 Hz and the pulse duration was 4 ms. Note that the fluctuations in the chilled supply temperature have a significant impact on the resonant frequency at these relatively low average RF power levels.

Table 1: Parameter Ranges Across Measured Data

| Parameter | Min | Max | Units |
|---|---|---|---|
| Wall Valve Setting | 0 | 99 | [% open] |
| Vane Valve Setting | 0 | 99 | [% open] |
| Cavity Field | 0 | 70 | [kV] |
| Wall Supply Temp. | 19.1 | 20.5 | [°C] |
| Vane Supply Temp. | 18.9 | 20.4 | [°C] |
| Wall Entrance Temp. | 19.8 | 22.8 | [°C] |
| Vane Entrance Temp. | 19.5 | 21.9 | [°C] |
| Resonant Frequency | 162.4403 | 162.4738 | [MHz] |
| Cave Temp. | 23.3 | 25.5 | [°C] |
| Cave Humidity | 19.1 | 36.6 | [%] |

## NEURAL NETWORK MODEL

### Input and Output

The inputs to the model are the temperature of the water entering each cooling sub-circuit (T01, TT01), the temperature of the water returning from the RFQ (T02, TT02), the two flow control valve read-backs, the ambient temperature and humidity, and a measure proportional to the power entering the cavity (given by the cavity field measurement and the duty factor). The output of the model is the predicted resonant frequency of the RFQ.

To exclude suspect measurements from training, target resonant frequency values are ignored when the cavity field drops below 0.5 kV (e.g. during multipactoring or sparking). However, as soon as the field recovers, the targets are used once again, thus including the thermal (and frequency) excursions caused by such interruptions.

### Network Architecture and Training

For this initial network, a simple feed-forward architecture with multiple previous time-steps embedded as inputs was selected. Initially, 30 minutes of previous system data were provided with a decaying sample interval. An initial topology and set of weights were obtained by conducting initial training and subsequently removing connections with small weights. The resultant network was then refined with further training. 10 networks with new initializations were trained using back-propagation in conjunction with scaled conjugate gradient optimization. Two hidden layers with 25 and 7 nodes in each layer respectively were used.

An approximate hyperbolic tangent activation function given by $g(x) = \frac{2}{(1+e^{-2x})} - 1$ was used for all nodes, except the output node, which used a linear activation function.

### Training, Validation, and Testing Sets

The validation data were interleaved with the training data (every-other sample). The testing data consists of a 2-D scan over vane valve settings and RF field amplitudes under a higher constant wall valve setting than was seen during training. The wall was set at 99% open for testing, whereas the highest prior value seen was 75% open.

### Performance

The best-performing network has a mean absolute prediction error of 346 Hz on the test set, 98 Hz on the validation set, and 116 Hz across all training, validation, and testing data. Figure 4 shows the predicted and measured resonant frequency for the scan shown in Figure 3.

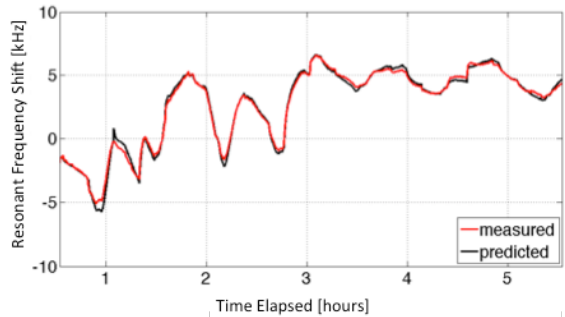

Figure 4: Measured and predicted resonant frequency values for the scan shown in Figure 3 (excluding data used for model initialization).

## CONCLUSION

We have created an initial neural network model that predicts the resonant frequency of the RFQ under changes in the cooling system and amount of RF heating. It performs sufficiently well for use in a model-based control routine. However, refinements could likely be made to the network structure to further improve performance. Data for CW operation will also need to be obtained, and training scans with finer granularity will be beneficial.

## ACKNOWLEDGMENT

The authors thank Maurice Ball and Jerzy Czaikowski for leading the water system design, as well as Tom Zuchnik and Dennis Nicklaus for their work on the water system instrumentation and data acquisition system. We also thank Ralph Pasquinelli, Dave Peterson, Ed Cullerton, and Philip Varghese for their work on the RF system. Many thanks also go to Curtis Baffes, Gennady Romanov, Jean-Paul Carneiro, Bruce Hanna, and Alexander Shemyakin for their many contributions to the design, installation, and commissioning of the RFQ. We also thank the team at LBNL, especially Andrew Lambert, Derun Li, and John Staples.